\documentclass[Afour,sagev,times]{sagej} %use this for SMMR style
\usepackage{moreverb,url}

\usepackage[colorlinks,bookmarksopen,bookmarksnumbered,citecolor=red,urlcolor=red]{hyperref}

\usepackage[utf8]{inputenc}
\usepackage{natbib}
\usepackage{fancyhdr}
\usepackage{float}
\usepackage{amssymb}
\usepackage{latexsym}
\usepackage{amsmath,bm,bbm}
\usepackage{amsfonts}
\usepackage{rotating}
\usepackage{graphicx} 
\usepackage{caption}
\usepackage{color,soul}
\usepackage{hyperref}
\hypersetup{colorlinks=true, linkcolor=blue, urlcolor=blue, citecolor=blue}
\usepackage{enumerate}
\usepackage{epsfig}
\usepackage{subfigure}
\usepackage{setspace}
\usepackage{appendix}
\usepackage{multirow,multicol}
\usepackage{booktabs,caption}
\usepackage[flushleft]{threeparttable}
\usepackage{dsfont}
\usepackage{url}  % Formatting web addresses
\usepackage{lipsum}

\newcommand\BibTeX{{\rmfamily B\kern-.05em \textsc{i\kern-.025em b}\kern-.08em
T\kern-.1667em\lower.7ex\hbox{E}\kern-.125emX}}

\def\volumeyear{2016}

\begin{document}

\runninghead{Hu et al.}

\title{Estimation of Causal Effects of Multiple Treatments in Observational Studies with a Binary Outcome}

\author{Liangyuan Hu\affilnum{1,2,3}, Chenyang Gu\affilnum{4}, Michael Lopez\affilnum{5}, Jiayi Ji \affilnum{1,2,3}, Juan Wisnivesky \affilnum{6} }

\affiliation{\affilnum{1}Department of Population Health Science and Policy,  Icahn School of Medicine, New York, USA\\
\affilnum{2}Institute for Health Care Delivery Science, Icahn School of Medicine, New York, USA\\
\affilnum{3}Tisch Cancer Institute, Icahn School of Medicine, New York, USA\\
\affilnum{4}Analysis Group, Inc., Los Angeles, USA\\
\affilnum{5}National Football League, New York, NY\\
\affilnum{6}Department of Medicine,Icahn School of Medicine, New York, NY\\
}

\corrauth{Liangyuan Hu, Center for Biostatistics, Department of Population Health Science and Policy, Ichan School of Medicine, New York, NY, 10029, USA.}

\email{liangyuan.hu@mssm.edu}

\begin{abstract}
There is a dearth of robust methods to estimate the causal effects of multiple treatments when the outcome is binary. This paper uses two unique sets of simulations to propose and evaluate the use of Bayesian Additive Regression Trees (BART) in such settings. First, we compare BART to several approaches that have been proposed for continuous outcomes, including inverse probability of treatment weighting (IPTW), targeted maximum likelihood estimator (TMLE), vector matching and regression adjustment. Results suggest that under conditions of  non-linearity and non-additivity of both the treatment assignment and outcome generating mechanisms, BART, TMLE and IPTW using generalized boosted models (GBM) provide better bias reduction and smaller root mean squared error. BART and TMLE provide more consistent 95 per cent CI coverage and better large-sample convergence property. Second, we supply BART with a strategy to identify a common support region for retaining inferential units and for avoiding extrapolating over areas of the covariate space where common support does not exist. BART retains more inferential units than the generalized propensity score based strategy, and shows lower bias, compared to TMLE or GBM, in a variety of scenarios differing by the degree of covariate overlap. A case study examining the effects of three surgical approaches for non-small cell lung cancer demonstrates the methods. 
\end{abstract}

\keywords{Causal inference; Generalized propensity score; Inverse probability of treatment weighting; Matching; Machine learning}

\maketitle

\section{Introduction}

\subsection{Motivating Research Question}
Lung cancer is the leading cause of cancer-related mortality worldwide and is estimated to have caused over 1.7 million deaths in 2018 \citep{lungcancer}. The most common type of lung cancer is non-small cell lung cancer (NSCLC), accounting for approximately 85$\%$ of all lung cancer cases \citep{molina2008non}. When feasible, NSCLC tumors are treated using surgical resection, which remains the  most effective option for a cure \citep{uramoto2014recurrence}. 

Open thoracotomy long stood as the standard surgical procedure for stage I-IIIA NSCLC tumors. However, open thoracotomy is associated with considerable postoperative complications and mortality, especially in the elderly \citep{scott2010comparison, whitson2008surgery}. Beginning in the late 1990s, two newer and less invasive techniques, video-assisted thoratic surgery (VATS) and, more recently, robotic-assisted surgery, were increasingly used \citep{park2012robotic, wisnivesky2011survival}. The adoption of VATS and robotic-assisted surgery seemed to signal that the newer procedures offer a clinical benefit relative to open resection \citep{yan2009systematic, toker2014robotic}. However, to our knowledge, no randomized controlled trials (RCTs) have been conducted to compare the effectiveness of these surgical procedures, in part due to difficulties in recruiting patients and high study costs. As a consequence, VATS and robotic-approaches were adopted into routine care without sufficient scrutiny \citep{park2012robotic, cajipe2012video}. 

In place of RCTs, large-scale population-based databases, such as the Surveillance, Epidemiology, and End Results (SEER)-Medicare database, provide research opportunities for comparative studies. The SEER-Medicare database comprises a large sample of patients who received each of the three surgical procedures and reflects patient outcomes in the real world setting, containing demographic and clinical information for Medicare beneficiaries with cancer in various United States regions \citep{warren2002overview}. However, in contrast to RCTs, the real-world adoption pattern of the three surgical approaches largely depends on the patients' sociodemographic and tumor characteristics, which may result in an unbalanced cohort with significant differences in the distributions of sociodemographic characteristics, comorbidities, cancer characteristics and diagnostic information across treatment groups \citep{veluswamy2017comparative}.

The research question poses several challenges for statistical analyses. First, in practice, statistical methods designed for a binary treatment are often used to account for underlying differences in patient characteristics to compare each pair of surgical procedures \citep{wisnivesky2010limited, cajipe2012video, park2012robotic}. Unfortunately, applications of these methods can lead to the comparisons of disparate patient subgroups, which may increase bias in treatment effect estimates \citep{lopez2017estimation}.  Second, common measures for comparative effectiveness are postoperative complications, which are binary outcomes. Thus, the treatment effects are typically based on the risk difference (RD), odds ratio (OD) or relative risk (RR) \citep{agresti2003categorical}, all of which make it less straightforward to obtain inference, relative to continuous outcomes \citep{austin2007performance,austin2008performance,austin2010performance}. Third, the robotic-assisted surgery is a new advanced technology that was just adopted into practice in recent years. As a result, the number of patients who are operated via this approach is smaller compared to the other two approaches, yielding unequal sample sizes across the treatment groups. Appropriate causal inference methods that can address these challenges are needed. 

\subsection{Overview of Methods for Causal Inference with Multiple Treatments} \label{sec:overview}

Recent years have seen a growing interest in the development of causal inference methods with multiple treatments using observational data. The theoretical work of Imbens\cite{imbens2000role} and Imai and Van \cite{imai2004causal} extended the propensity score framework in the setting with a binary treatment \citep{rosenbaum1983central} to the general treatment setting. Subsequently, methods designed for a binary treatment have been reformulated to accommodate multiple treatments, including regression adjustment (RA) \citep{linden2016estimating}, inverse probability of treatment weighting (IPTW) \citep{feng2012generalized, mccaffrey2013tutorial}, and vector matching (VM) \citep{lopez2017estimation}. Lopez and Gutman \cite{lopez2017estimation} provide a comprehensive review of current methods for multiple treatments. These methods focus on continous outcomes.

RA \citep{rubin1973use, rubin1979using, linden2016estimating}, also known as model-based imputation \citep{imbens2015causal} uses a regression model to impute missing outcomes, estimating what would have happened to a specific unit had this unit received the treatment to which it was not exposed. The causal estimand of interest can be estimated by contrasting the imputed potential outcomes between treatment groups. The critical part of this method is the specification of the functional form of the regression model. With a low-dimensional set of pre-treatment covariates, it is relatively easy to specify a flexible functional form for the regression model. If there are many pre-treatment covariates, however, such a specification is more difficult, and possible misspecification of the regression model could bias the estimate of treatment effects. RA also heavily relies on extrapolation for estimation when the covariate distributions between treatment groups are far apart \citep{imbens2015causal}.

IPTW \citep{imbens2000role, feng2012generalized, mccaffrey2013tutorial} methods attempt to obtain an unbiased estimator for treatment effects in a way akin to how weighting by the inverse of the selection probability adjusts for unbalances in sampling pools, introduced by Horvitz and Thompson \cite{horvitz1952generalization} in survey research. A challenge with IPTW is that treated units with low generalized propensity scores that are close to zero can result in extreme weights, which may yield erratic causal estimates with large sample variances \citep{little1988missing, kang2007demystifying}. This issue is increasingly likely as the number of treatments increases \citep{lopez2017estimation}. An alternative method is to use trimmed or truncated weights, in which weights that exceed a specified threshold are each set to that threshold \citep{cole2008constructing, lee2011weight}. The threshold is often based on quantiles of the distribution of the weights (e.g., the 1st and 99th percentiles). 

Alternatives to estimate generalized propensity scores in the IPTW framework include generalized boosted models\cite{mccaffrey2013tutorial} (GBM) and Super Learner\cite{van2007super, rose2019double} (SL). GBM grows multiple regression trees to capture complex and nonlinear relationships between treatment assignment and pre-treatment variables. The estimation procedure can be tuned to find the generalized propensity score model producing the best covariate balance between treatment groups. This feature of GBM should help alleviate extreme weights and improve the estimation of causal effects \citep{mccaffrey2013tutorial}. However, the algorithm can be computationally intensive, and the robust procedure for estimating the variances of the effect estimates is not guaranteed to result in proper confidence intervals. SL uses ensemble of machine learning approaches including regression, ridge regression, and classification trees, to estimate a weight for each treatment. There is no guarantee that these probabilities sum to 1, and it is common to normalize weights accordingly. To limit extreme weights, Rose and Normand\cite{rose2019double} use a lower bound of 0.025 for each probability.

Targeted Maximum Likelihood Estimation \cite{schuler2017targeted, rose2019double} (TMLE) is a doubly robust approach that combines outcome estimation, IPTW estimation, and a targeting step to optimize the parameter of interest with respect to bias/variance. Rose and Normand \cite{rose2019double} implements TMLE by estimating both generalized propensity scores and a binary outcome using SL. For obtaining variance estimates, Rose and Normand \cite{rose2019double} uses influence curves, though bootstrapping is also suggested. The use of TMLE  has, to the best of our knowledge, not been deeply vetted for multiple treatment options by using simulations.

Lopez and Gutman\cite{lopez2017estimation} proposed the VM algorithm, which can match units with similar vector of generalized propensity scores. VM is designed to replicate a multi-arm randomized trial by generating sets of units that are roughly equivalent on measured pre-treatment covariates. VM obtains matched sets using a combination of $k$-means clustering and one-to-one matching with replacement within each cluster strata. Simulations demonstrated that, relative to IPTW with the generalized propensity scores estimated using multinomial logistic regression, and to generalizations of tools designed for binary treatment, VM yielded lower bias in the covariates' distributions between different treatment groups, while retaining most of eligible units that received the reference treatment \citep{lopez2017estimation}. However, the authors acknowledge that there is a lack of guidance regarding the estimation of the sampling variance, and this is an area for further statistical research. 
%\cite{scotina2018matchingA} extends the idea of VM to matching with Mahalanobis distance and the use of fuzzy matching. 

Before describing BART -- one tool that we think is equipped to handle the complexity of causal inference with multiple treatments --  it is worth explicating on why one approach that, although intuitive and easy to perform, is not recommended: a series of binary comparisons (SBC). To wit, grouping subjects into separate sub-populations, each with two treatments, and then using approaches designed for binary treatment is an approach often used in practice \citep{lopez2017estimation}. However, SBC can (i) lead to non-transitive causal estimates, (ii) increase bias, and (iii) leave it unclear which treatment is optimal, all of which make it inappropriate for causal inference when there are more than two treatments \citep{lopez2017estimation}.

\subsection{Bayesian Additive Regression Trees for Causal Inference}

While the advanced regression and propensity score-based techniques described above were created for causal inference with multiple treatments, these methods were developed with continuous outcomes in mind, and they have been less studied in the context of both a binary outcome and multiple treatments.

In recent years, Bayesian Additive Regression Trees (BART) \citep{chipman2007bayesian, chipman2010bart}, a nonparametric modeling tool, has become more popular in causal settings. Hill\cite{hill2011bayesian} proposed the use of BART for causal inference with a binary treatment and a continuous outcome. Hill\cite{hill2011bayesian} and Hill and Su\cite{hill2013assessing} used simulations to show that, in scenarios where there are nonlinearities in the response surface and the treatment assignment mechanism, BART generates more accurate estimates of average treatment effects compared to various matching and weighting techniques, and comparable estimates in linear settings. 

BART boasts several advantages for causal inference with a binary treatment \citep{hill2011bayesian, dorie2016flexible}. First, BART allows for an extremely flexible functional form. Second, BART avoids ambiguity with respect to covariate balance diagnostics required by propensity score based approaches. Third, BART generates coherent uncertainty intervals for treatment effect estimates from the posterior samples in contrast to propensity score matching and subclassification, for which there is lack of agreement regarding appropriate interval estimation \citep{imai2004causal,hill2011bayesian}. Finally, BART is easy to implement and requires less researcher programming expertise. However, like any methods that do not first discard units that fall out of  areas of the covariate space where common support does not exist, one vulnerability of BART is that there is no mechanism to prevent it from extrapolating over these areas. 

We surmise that the strengths of BART are transferable to the multiple treatment setting. In the sections that follow, we conduct two sets of simulations to investigate the operating characteristics of BART for estimating the causal effects of multiple treatments on a binary outcome, and compare BART to the existing methods discussed previously. We further supply BART with a strategy to identify a common support region and compare it to the propensity score based strategy with respect to the proportion of units retained for inference and the accuracy of treatment effect estimates based on the retained inferential units. We subsequently apply the methods examined to analyze a large data set on stage I-IIIA NSCLC patients, drawn from the SEER-Medicare registry, and estimate the comparative effect of robotic-assisted surgery versus VATS and open thoracotomy on postoperative outcomes. 

%This paper is organized as follows: Section 2 describes the potential outcomes framework for multiple treatments. Section 3 presents the use of BART to estimate causal effects in the multiple treatment setting. Section 4 describes the importance of common support in causal inference and proposes a strategy for BART to identify a common support region for inferential units. Section 5 details the design and results of two Monte Carlo Simulations. Section 6 provides a case study examining the effects of three surgical approaches for NSCLC patients.  Conclusions and discussions are provided in Section 7. Simulation code and a walk-through example using BART are provided at (site available upon publication). 

%%%%%%%%%%%%%%%%%%%%%%%%%%%%%%%%%%%%%%%%%%%
\section{Potential Outcomes Framework for Multiple Treatments}

\subsection{Notation and Assumptions}
Our notation is based on the potential outcomes framework, which was originally proposed by Neyman \cite{Neyman1990application} in the context of randomization-based inference in experiments. Potential outcomes were generalized to observational studies and Bayesian analysis by Rubin \cite{rubin1974estimating, rubin1977assignment, rubin1978bayesian}, in what is now known as the Rubin Causal Model (RCM) \citep{holland1986statistics}. 

Consider a sample of $N$ units, indexed by $i = 1,\ldots,N$, drawn randomly from a target population, which comprises individuals in a study designed to evaluate the effect of a treatment $W$ on some outcome $Y$. Each unit is exposed to one of total $Z$ possible treatments; that is, $W_i = w$ if individual $i$ was observed under treatment $w$, where $w \in \mathcal{W} = \{1,2,\ldots,Z\}$. The number of units receiving treatment $w$ is $n_{w}$, where $\sum_{w=1}^Z n_{w} = N$. For each unit $i$, there is a vector of pre-treatment covariates, $\bm{X}_i$, that are not affected by $W_i$. Let $Y_i$ be the observed outcome of the $i$th unit given the assigned treatment, and $\{Y_i(1),\ldots,Y_i(Z)\}$ the potential outcomes for the $i$th unit under each treatment of $\mathcal{W}$. For each unit, at most one of the potential outcomes is observed (the one corresponding to the treatment to which the unit is exposed). All other potential outcomes are missing, which is known as the fundamental problem of causal inference \citep{holland1986statistics}. Let $r(w,\bm{X}_i)$ be the generalized propensity score (GPS), which is defined as the probability of receiving treatment $w$ given pre-treatment covariates, that is, $r(w,\bm{X}_i) = Pr(W_i = w | \bm{X}_i)$, for $\forall$ $w \in \{1,\ldots,Z\}$ \citep{imbens2000role, imai2004causal}. This definition extends the propensity score \citep{rosenbaum1983central} from a binary treatment setting to the multiple treatment setting, in which conditioning must be done on a vector of GPSs, defined as $\bm{R}(\bm{X}_i) = (r(1,\mathbf{X}_i),\ldots, r(Z,\bm{X}_i))$, or a function of $\bm{R}(\bm{X}_i)$ \citep{imai2004causal}. In addition, we define the response surface as $f(w,\bm{X}_i) \equiv E[Y_i(w)|\bm{X}_i]$, for $w \in \{1,\ldots,Z\}$.

In general, causal effects are not identifiable without further assumptions because only one of the potential outcomes is observed for every unit. We make the following identifying assumptions:
\begin{enumerate}
    \item The stable unit treatment value assumption (SUTVA) \citep{rubin1980randomization}, that is, no interference between units and no different versions of a treatment.
    
    \item The positivity or sufficient overlap assumption; that is, $0 < p(W_i|Y_i(1),\ldots,Y_i(Z),\bm{X}_i) < 1$, $\forall \; W_i \in \{1,\ldots,Z\}$, which implies that there are no values of pre-treatment covariates that could occur only among units receiving one of the treatments.
    
    \item The treatment assignment is unconfounded; that is, $p(W_i|Y_i(1),\ldots,Y_i(Z),\bm{X}_i) = p(W_i|\bm{X}_i)$, $\forall \; W_i \in \{1,\ldots,Z\}$, which implies that the set of observed pre-treatment covariates, $\bm{X}_i$, is sufficiently rich such that it includes all variables directly influencing both $W_i$ and $Y_i$; in other words, there is no unmeasured confounding.
\end{enumerate}

Under the unconfoundedness assumption, for any treatment $w$ and pre-treatment covariates $\bm{X}_i$,
\begin{equation}
f(w,\bm{X}_i) = E[Y_i(w) | W_i=w, \bm{X}_i] = E[Y_i | W_i=w, \bm{X}_i],
\end{equation}
where the second identity is the conditional mean function of the observed outcomes.

\subsection{Definition of Causal Effects}
Causal effects are summarized by estimands, which are functions of the unit-level potential outcomes on a common set of units \citep{rubin1974estimating, rubin1978bayesian}. For dichotomous outcomes, causal estimands can be the RD, OD or RR. For purposes of illustration, we use RD in this paper. 

Following Lopez and Gutman\cite{lopez2017estimation}, we provide a broad definition of the causal risk difference that may be of interest with multiple treatments. Define $s_1$ and $s_2$ as two subgroups of treatments such that $s_1,s_2 \subset \mathcal{W}$ and $s_1 \cap s_2 = \emptyset$. Next, let $|s_1|$ and $|s_2|$ be the cardinality of $s_1$ and $s_2$, respectively. Two commonly used causal estimands are the average treatment effect (ATE), $ATE_{s_1,s_2}$, and the average treatment effect among those receiving $s_1$, $ATT_{s_1|s_1,s_2}$,
where 
\begin{equation}
\begin{split}
\label{eq:pop_est}
ATE_{s_1,s_2} &= E \bigg{[} \frac{\sum_{w \in s_1} Y_i(w)}{|s_1|} - \frac{\sum_{w' \in s_2} Y_i(w')}{|s_2|} \bigg{]},\\
ATT_{s_1|s_1,s_2} &= E \bigg{[} \frac{\sum_{w \in s_1} Y_i(w)}{|s_1|} - \frac{\sum_{w' \in s_2} Y_i(w')}{|s_2|} \bigg{|} W_i \in s_1 \bigg{]}.
\end{split}
\end{equation}

%\begin{equation}
%\resizebox{0.5\textwidth}{!}{
%\begin{split}
%\label{eq:pop_est}
%ATE_{s_1,s_2} &= E \bigg{[} \frac{\sum_{w \in s_1} Y_i(w)}{|s_1|} - \frac{\sum_{w' \in %s_2} Y_i(w')}{|s_2|} \bigg{]},\\
%ATT_{s1|s_1,s_2} &= E \bigg{[} \frac{\sum_{w \in s_1} Y_i(w)}{|s_1|} - \frac{\sum_{w' \in %s_2} Y_i(w')}{|s_2|} \bigg{|} W_i \in s_1 \bigg{]}.
%\end{split}}
%\end{equation}

In~\eqref{eq:pop_est}, the expectation is over all units, $i = 1,\ldots,N$, and the summation is over the potential outcomes of a specific unit. Another set of causal estimands are the conditional treatment effects, given pre-treatment covariates $\bm{X}_i$,
\begin{equation}
%\resizebox{0.5\textwidth}{!}{
\begin{split}
 CATE_{s_1,s_2} 
 &= E \bigg{[} \frac{\sum_{w \in s_1} Y_i(w)}{|s_1|} - \frac{\sum_{w' \in s_2} Y_i(w')}{|s_2|} \bigg{|} \bm{X}_i; \theta_{Y|X} \bigg{]}\\
CATT_{s_1|s_1,s_2} 
&= E \bigg{[} \frac{\sum_{w \in s_1} Y_i(w)}{|s_1|} - \frac{\sum_{w' \in s_2} Y_i(w')}{|s_2|} \bigg{|} W_i \in s_1, \bm{X}_i; \theta_{Y|X} \bigg{]}
\end{split}
%}
\end{equation}
where the parameters $\theta_{Y|X}$ and $\theta_X$ index the conditional distribution $p(Y|\mathbf{X})$ and distribution $p(\mathbf{X})$, respectively. 

Causal inference methods via modeling the response surfaces (e.g., BART and RA) arrive at the population or sample marginal treatment effects by integrating the conditional effects over the distribution of $\bm{X}_i$ \citep{ding2018causal}. In most cases, however, it is difficult to model the possibly multi-dimensional $\bm{X}_i$. We can obtain the marginal effects by averaging the treatment effects conditional on the observed values of the covariates over the empirical distribution of $\{\bm{X}_i\}_{i=1}^N$, 
\begin{equation}
\begin{split}
ATE_{s_1,s_2} &= \int CATE_{s_1,s_2}(\bm{X}, \theta_{Y|X}) dF_X(\bm{X};\theta_X)\\
ATT_{s_1|s_1,s_2} &= \int CATT_{s1|s_1,s_2}(\bm{X}, \theta_{Y|X}) dF_X(\bm{X};\theta_X),
\end{split}
\end{equation}

In our motivating example, one of the research questions of interest is to compare the effectiveness of a newer minimally-invasive procedure (i.e., robotic-assisted surgery) versus the existing surgical procedures (e.g., VATS) in the overall population, or among those patients who received robotic-assisted surgery. The corresponding target causal estimands are defined as
\begin{equation}
\begin{split}
   ATE_{1,2} &= \int E[Y_i(1) - Y_i(2) | \bm{X}_i; \theta_{Y|X}] dF_X(\bm{X};\theta_X)\\
   ATT_{1|1,2} &= \int E[Y_i(1) - Y_i(2) | W_i = 1, \bm{X}_i; \theta_{Y|X}] dF_X(\bm{X};\theta_X).
 \end{split}   
\end{equation}

\subsection{Treatment effects using BART}

Under the identifying assumptions, treatment effects such as $ATT_{1|1,2}$ can be estimated by contrasting the imputed potential outcomes between robotic-assisted surgery and VATS groups among those patients who received robotic-assisted surgery, predicted from the estimates of the respective response surface models. In principle, any method that can flexibly estimate $f(w,\bm{X}_i)$ could be used to predict the potential outcomes. \cite{chipman2007bayesian, chipman2010bart} demonstrated that BART has important advantages as a predictive algorithm over alternative methods in the machine learning literature such as classification and regression trees \citep{breiman1984classification}, boosting \citep{freund1997desicion} and random forests \citep{breiman2001random}, in particular with regard to choosing tuning parameters and generating coherent uncertainty intervals. 

BART is a Bayesian ensemble method that models the mean outcome given predictors by a sum of trees. For a binary outcome, the BART model can be expressed using the probit model setup as:
\begin{eqnarray}
f(w,\bm{X}_i) &=& E(Y_i | W_i = w, \bm{X}_i) = \Phi \bigg{\{} \sum_{j=1}^J g_j(w, \bm{X}_i; T_j, M_j) \bigg{\}},
\end{eqnarray}
where $\Phi$ is the the standard normal c.d.f., each $(T_j, M_j)$ denotes a single subtree model in which $T_j$ denotes the regression tree and $M_j$ is a set of parameter values associated with the terminal nodes of the $j$th regression tree, $g_j(w,\bm{X}_i)$ represents the mean assigned to the node in the $j$th regression tree associated with covariate value $\bm{X}_i$ and treatment level $w$, and the number of regression trees $J$ is considered to be fixed and known. The details of the specification of prior distribution and the choice of hyper-parameters can be found in Chipman et al.\cite {chipman2010bart}. Sampling from the posterior distributions proceed via a Bayesian backfitting MCMC algorithm \cite{chipman2010bart}. A total of $L$ Markov Chain Monte Carlo (MCMC) samples of model parameters, $(T_j,M_j)$, are drawn from their posterior distribution. For each of $L$ draws, we predict the potential outcomes for each unit and the relevant treatment level. The causal estimand of interest can be estimated by contrasting the imputed potential outcomes between treatment groups. For example, $ATT_{1|1,2}$ can be estimated as follows:
\begin{equation}
\begin{split}
\widehat{ATT}_{1|1, 2} &= (n_{1}L)^{-1} \sum_{l=1}^L \sum_{i:W_i=1}  \bigg{\{}  f^l(1, \bm{X}_i) - f^l(2, \bm{X}_i)   \bigg{\}}\\
&= (n_{1}L)^{-1} \sum_{l=1}^L \sum_{i:W_i=1}  \bigg{\{} \Phi \bigg{[} \sum_{j=1}^J g_j(1, \bm{X}_i; T^l_j, M^l_j) \bigg{]} - \Phi \bigg{[} \sum_{j=1}^J g_j(2, \bm{X}_i; T^l_j, M^l_j) \bigg{]}   \bigg{\}},
\end{split}
\end{equation}
where $(T^l_j,M^l_j)$ are the $l$th draw from the posterior distribution of $(T_j,M_j)$. We can obtain obtain the point and interval estimates of the treatment effect directly using the summary of posterior samples.

\subsection{Common Support}

Because problems can arise when drawing inference to regions of the covariate space where there are insufficient number of units in all treatment groups, propensity score based methods are typically equipped with strategies for defining a common support region. For BART, there is no such a mechanism to prevent it from extrapolating over areas where a common support does not exist. 

For a binary treatment, one strategy is to discard units that fall beyond the range of the propensity score \citep{dehejia2002propensity, morgan2006matching}. Hill and Su\cite{hill2013assessing} argue that these strategies typically ignore the information embedded in the response variable, and propose alternative discarding rules. Illustrative examples with one or two predictors were used to compare the two types of discarding strategies and their implications on estimation of the causal effects and the proportion of inferential units retained. Advantages of BART over the propensity score approach manifest in examples where there is lack of common support for variables only predictive of treatment but not of the outcome or the treatment mechanism is more difficult to model. However, in practice, identifying common support is often required for a high-dimensional covariate space. In addition, the two types of strategies have not been compared in the multiple treatment setting. 

%, and compare it to the GPS-based strategy, employed by VM. To assess common support frameworks, we compare the proportion of units retained for inference and the accuracy of the causal effect estimates after applying the discarding rules. 

To address these limitations, we propose a strategy for BART to define both a common support region and the corresponding discarding rules. Whereas Hill and Su\cite{hill2013assessing} uses a common support for binary treatment using the \emph{1 sd rule}, our empirical simulations suggest this rule may be too relaxed in the setting of three or more treatment groups. We use a sharper cutoff and identify a common support as follows. We discard any unit $i$, with $W_i = w$, for which $s_i^{f_{w'}} > \text{max}_j \{s_j^{f_w} \}$ , $\forall j: W_j = w$, where $s_j^{f_w}$ and $s_j^{f_{w'}}$ denote the standard deviation of the posterior distribution of the potential outcomes under treatment $W = w$ and $W=w'$, respectively, for a given unit $j$.

For multiple treatments with $Z=3$, when estimating the ATT of treatment $W = 2$ and $W=3$ among those treated with $W = 1$, we discard for unit $i$ with $W_i = 1$, if 
\begin{equation}
\label{eq:1sd}
\begin{split}
s_i^{f_2} &> \text{max}_j \{s_j^{f_1} \}  ,  \hspace{5mm} \text{and}\\
s_i^{f_3} &> \text{max}_j \{s_j^{f_1} \}.  
\end{split}
\end{equation}
When estimating the ATE, we apply the discarding rule in~\eqref{eq:1sd} to each treatment group. 

There is likewise a lack of consensus for defining a common support region with GPS-based approaches. For matching using the GPS, \cite{lopez2017estimation} propose a rectangular support region. Let $r(w, \bm{X})$ denote the treatment assignment probability for $w$, and let $r(w, \bm{X} | W = w')$ represent treatment assignment probability for $w$ among those who received treatment $w'$. A rectangular common support region can be defined as follows with $Z = 3$. For any $w, w' \in \mathcal{W} = \{1,2,3\}$, 
\begin{equation}
\label{eq:gpscomsup}
\begin{split}
r(w, \bm{X})^{low} &= \max\{\min(r(w, \bm{X} | W  = 1)),  \min (r(w, \bm{X} | W  = 2)),  \min(r(w, \bm{X} | W  = 3))\} \\
r(w, \bm{X})^{high} &= \min\{\max(r(w, \bm{X} | W  = 1)), \max(r(w, \bm{X} | W  = 2)),  \max(r(w, \bm{X} | W  = 3)) \}
\end{split}
\end{equation}

For weighting methods, techniques such as trimming \citep{lee2011weight} or stabilizing (more useful for time-varying confounding, see Hu et al. \cite{hu2017modeling}, Hu and Hogan \cite{hu2019causal} and Hern{\'a}n and Robins \cite{hernancausal2019}) are frequently used in place of a common support.  However, the lack of common support in the covariate space may lead to extreme weights and unstable IPTW estimators.

%%%%%%%%%%%%%%%%%%%%%%%%%%%%%%%%%%%%%%%%%%%
\section{Simulation Studies}\label{sec:sim}

%In this section, we follow the simulation design described in \citet{hill2013assessing} to explore evidence regarding the performance of BART and alternatives methods described in Section~\ref{sec:approaches} for estimation of causal effects with multiple treatments. 

\subsection{Design and implementation}

We conduct expansive simulations in order to better understand how BART will work in complex causal settings. Our first set of simulations, Simulation 1, contrasts BART with other approaches, while our second set, Simulation 2, looks into the role that covariate overlap plays in inferences with multiple treatments. 

The design of both simulations mimics the range of scenarios that are representative of the data structure in the SEER-Medicare registry. Three treatment levels ($Z=3$) are used throughout, with pairwise ATTs of risk difference are our outcome of interest. True treatment effects are computed based on a simulated super-population of size 100,000. We replicated each of the scenarios described below 200 times within sub-populations of the superpopulation. In Simulation 1, we began with the comparisons of 10 methods: 1) RA; 2) IPTW with weights estimated using multinomial logistic regression (IPTW-MLR); 3) IPTW with weights estimated using generalized boosted models (IPTW-GBM); 4) IPTW with weights estimated using super learner (IPTW-SL); 5) IPTW-MLR with trimmed weights; 6) IPTW-GBM with trimmed weights; 7) IPTW-SL with trimmed weights; 8) VM;  9) TMLE; 10) BART. We used VM to only estimate the ATT effects as the algorithm for estimating the ATE has not been fully developed, and implemented TMLE to only estimate the ATE effects as we are not aware of any implementation of TMLE for the estimation of ATT effects for multiple treatment options. In simulation 2, only BART, TMLE and IPTW-GBM, the top performing methods in Simulation 1, were further examined.

We implemented the methods as follows. For RA, we first fit a Bayesian logistic regression model with main effects of all confounders using the \texttt{bayesglm()} function in the \texttt{arm} package in \verb+R+. We then drew a total of 1,000 MCMC samples of regression coefficients from their posterior distributions and predicted the potential outcomes for each unit and relevant treatment group. When implementing IPTW, we estimated GPSs by including each confounder additively to a multinomial logistic regression model, a generalized boosted model, and a super learner model respectively. The stopping rule for the optimal iteration of GBM was based on maximum of absolute standardized bias, which compares  the distributions of the covariates between treatment groups \citep{mccaffrey2013tutorial}. We implemented SL using 
the \texttt{weightit()} function in the \verb+R+ package \texttt{WeightIt} for multinomial treatment and included three algorithms: main terms regression, generalized additive model, and support vector. The treatment probabilities are normalized to sum to one. The weights -- inverse of the GPSs -- were then trimmed at 5\% and 95\% to generate trimmed IPTW estimators. GPSs for VM were estimated using multinomial logistic regression with main effects of all confounders. We used a combination of $k$-means clustering with $k$ = 5 subclasses and one-to-one matching with replacement and a caliper of 0.25 to ensure that the matched cohort is relatively similar in terms of the distributions of the confounders. We used the \verb+R+ package \texttt{tmle} to implement TMLE as described in Rose and Normand\cite{rose2019double}. We used SL to estimate each treatment probability and bound them from below to 0.025. Applying BART to the simulation datasets, we used the default priors associated with the \texttt{bart()} function available in the \texttt{BART} package in \verb+R+. For each BART fit, we allowed the maximum number of trees in the sum to be 100. To ensure the convergence of the MCMC in BART, we let the algorithm run for 5000 iterations with the first 3000 considered as burn-in.

To judge the appropriateness of each technique, we use mean absolute bias (MAB), root mean squared error (RMSE) and coverage probability (CP). In addition, we examine the large-sample convergence property of each method. 

\subsubsection{Simulation 1: which causal approach yields the lowest bias and RMSE?}

We compare each of the 10 approaches across a combination of two design factors: the study sample size (i.e., the total number of units) and the ratio of units in the treatment groups. We varied the two factors in three scenarios: 1) 1200 with a 1:1:1 ratio, 2) 4000 with a 1:5:4 ratio,  and 3) 11,600 with a 1:15:13 ratio to represent equal, moderately unequal and highly unequal sample sizes across treatment groups.  The relatively small sample size (400) in the first group -- which will be used as the reference group of the ATT effects -- and the scenario of highly unequal sample sizes mimic the SEER-Medicare data in the motivating study. 

We considered 10 confounders with five continuous variables and five categorical variables. We assumed that both the treatment assignment mechanism and the response surfaces are nonlinear models of the confounders, as a realistic representation of the application data. Specifically, the treatment assignment follows a multinomial logistic regression model,  
\begin{equation}
\label{eq:trtmod}
\begin{split}
\ln  \dfrac{P(W=1)}{P(W=3)} &= \alpha_1 + \mathbf{X}\xi_1^L + \mathbf{Q}\xi_1^{NL}\\
\ln  \dfrac{P(W=2)}{P(W=3)} &= \alpha_2 + \mathbf{X}\xi_2^L + \mathbf{Q}\xi_2^{NL}
\end{split}
\end{equation}
where $\mathbf{Q}$ denotes the nonlinear transformations and higher-order terms of the predictors $\mathbf{X}$, $\xi_1^L$ and  $\xi_2^{L}$ are vectors of coefficients for the untransformed versions of the predictors $\mathbf{X}$ and $\xi_1^{NL}$ and  $\xi_2^{NL}$ for the transformed versions of the predictors captured in $\mathbf{Q}$. 
The intercepts, $\alpha_1$, $\alpha_2$, were specified to create the corresponding ratio of units in three treatment groups in each scenario. We generated three sets of parallel response surfaces as follows:
\begin{equation}
\label{eq:po}
\begin{split}
E[Y(1) | \mathbf{X}]& = \text{logit}^{-1}  \{ \tau_1+  \mathbf{X}\gamma^{L} + \mathbf{Q} \gamma^{NL} \} \\
E[Y(2) | \mathbf{X}]& = \text{logit}^{-1}  \{   \tau_2+ \mathbf{X}\gamma^{L} + \mathbf{Q} \gamma^{NL}  \}\\
E[Y(3) | \mathbf{X}]& =  \text{logit}^{-1}  \{ \tau_3+\mathbf{X}\gamma^{L} + \mathbf{Q} \gamma^{NL} \}
\end{split}
\end{equation}
where regression coefficients ($\tau_1, \tau_2, \tau_3$, $\gamma^L$, and $\gamma^{NL}$) were chosen so that the prevalence rates in the treatment groups were similar as the rates of respiratory complications observed in the SEER-Medicare data (see Table~\ref{tab:eventrate}). By generating nonparallel response surfaces across treatment groups, we can induce heterogeneous treatment effects. This topic warrants a stand-alone research and is beyond the scope of this article. Details of model specification in Equation~\eqref{eq:trtmod} and~\eqref{eq:po} are given in Table S1 of Supplementary Materials.
The observed outcome $Y$ is related to the potential outcome $Y(w)$ via 
$Y_i = \sum_{w \in\{w_1,w_2,\ldots,w_{Z}\}} Y_i(w) I(W_i = w)$. 

\subsubsection{Simulation 2: How do levels of covariate overlap impact causal estimates}

Only BART, TMLE and IPTW-GBM, the top performing methods in Simulation 1, are used in Simulation 2, which more deeply examines the impact of covariate overlap. 

We generated datasets following the simulation configuration of scenario 3 in Simulation 1, including the total sample size, the ratio of units, the number of continuous and categorical confounders and the response surface models, to mimic the SEER-Medicare dataset. To create varying covariate overlap that are ``measurable'' in degrees, we generate the treatment variable and covariate distribution as follows. 

Three levels of covariate overlap were designed: 1) \emph{weak} -- there is lack of overlap in the covariate space defined by all 10 confounders, 2) \emph{strong} -- there is strong overlap with respect to each of the 10 confounders, and 3) \emph{moderate} -- the five categorical variables had sufficient overlap as in the \emph{strong} scenario and overlap is lacking for the five continuous variables. Two configurations were examined in the \emph{moderate} scenario.  All of the five continuous variables or only two of them were included in the response surface models, resulting in one configuration where overlap was lacking for a variable that was a true confounder and another configuration when overlap was lacking for a variable that was not predictive of the response surface (therefore not a true confounder). This simulation is designed to  make it difficult for any method to successfully estimate the true treatment effect, as both the treatment assignment and the outcome are difficult to model. We simulated datasets for each scenario as follows.
 
 \begin{itemize}
\item \emph{Weak}. We assumed that the treatment variable $W$ followed a multinomial distribution, $W \sim \text{Multinomial} (N, p_1, p_2, p_3)$, and generated the treatment assignment by setting $N = 11600$, $p_1 = .03$,  $p_2 = .52$ and $p_3 = .45$. The covariates were generated from the distributions conditional on treatment assignment to create sufficient or lack of overlap. The continuous variables were generated independently from $X_j | W = 1 \sim N(-1,1)$, $X_j | W = 2 \sim N(1,1)$,   $X_j | W = 3 \sim N(3,1)$ for $j = 1, \ldots, 5$. The categorical variables were generated independently from $X_j | W= 1 \sim \text{Multinomial} (N, .3, .3, .4)$,  $X_j | W= 2 \sim \text{Multinomial} (N, .6, .2, .2)$, $X_j | W= 3 \sim \text{Multinomial} (N, .8, .1, .1)$, for $j = 6, \ldots, 10$. The potential outcomes of each treatment group were drawn from the response surface models~\eqref{eq:po}, with all of the 10 covariates included (i.e., all covariates are true confounders). Under this scenario, lack of overlap was designed for each of the 10 confounders. 

\item \emph{Strong}.  The treatment variable $W$ was generated in the same way as in the \emph{weak} scenario. We created strong covariate overlap by generating similar distributions of the covariates across the treatment groups for all 10 confounders $X_1 - X_{10}$. Specifically, we assumed $X_j | W \sim N(.05W,1-0.05W)$ for $j = 1, \ldots, 5$, and $X_j | W \sim \text{Multinomial} (N, .3 - .001W, .3 +.001W, .4)$ for $j = 6, \ldots, 10$. 

\item \emph{Moderate}. We generated five categorical confounders $X_6 - X_{10}$ with strong overlap, and lack of overlap for five continuous variables $X_1 - X_{5}$. We distinguished the situation where overlap is lacking for a variable that is not predictive of the outcome (\emph{moderate} I) and the situation when it is lacking for a true confounder (\emph{moderate} II). Specifically, we assumed that $X_j | W = 1 \sim N(-0.5,1)$, $X_j | W = 2 \sim N(1,1)$,   $X_j | W = 3 \sim N(2.5,1)$ for $j = 1, \ldots, 5$ and $X_j | W \sim \text{Multinomial} (N, .3 - .001W, .3 +.001W, .4)$ for $j = 6, \ldots, 10$. In \emph{moderate} I, the response surface models only included covariates $X_1, X_5, X_6 - X_{10}$, thus,  $X_2, X_3$ and $X_4$ that defined a covariate area in which the lack of overlap ocurred are non-confounders. In \emph{moderate} II, covariates $X_1 - X_{10}$ were all included in the response surface model, inducing lack of overlap in five true confounders. 
\end{itemize}

\begin{figure*}[ht]
\begin{center}
\subfigure[Weak Covariate Overlap]{\label{fig:overlap-a}\includegraphics[scale=0.18]{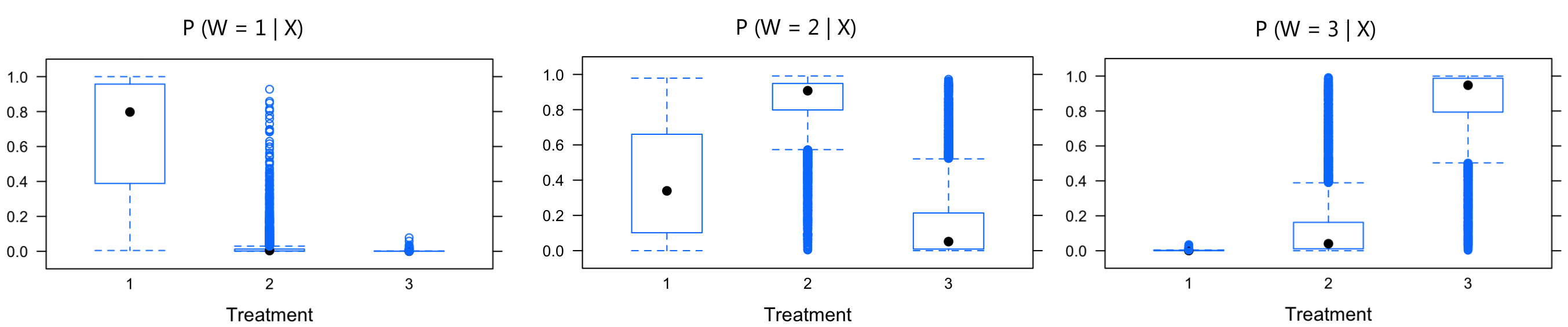}}
\subfigure[Moderate Covariate Overlap]{\label{fig:overlap-b}\includegraphics[scale=0.18]{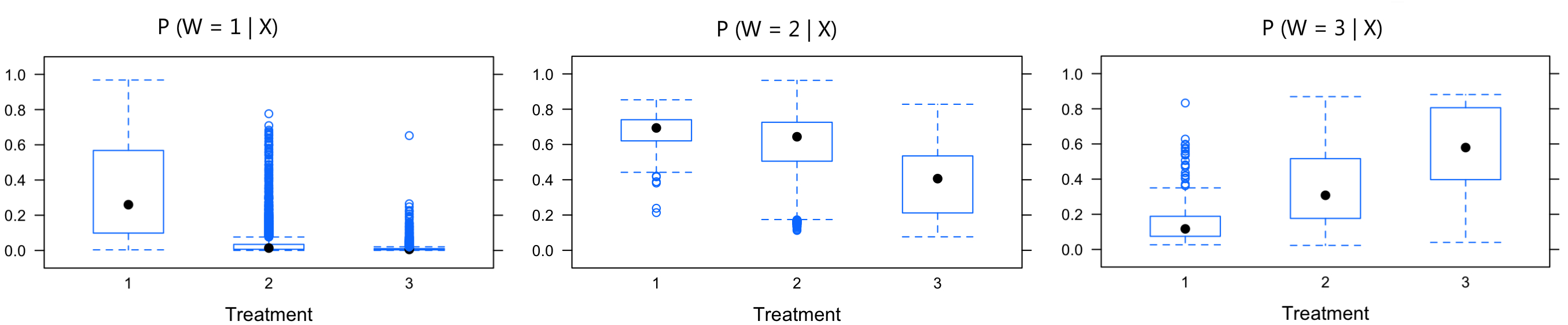}}
\subfigure[Strong Covariate Overlap]{\label{fig:overlap-c}\includegraphics[scale=0.18]{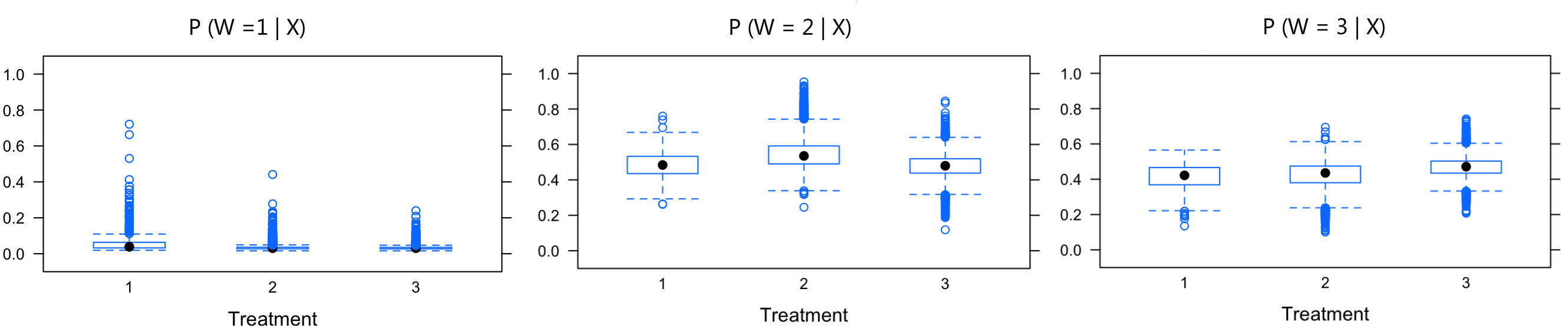}}
\caption{Overlap assessment for the scenarios of weak, moderate and strong covariate overlap. Each panel presents boxplots by treatment group of the estimated generalized propensity scores for one of the treatments, $P(W_i = w| X)$, $w \in \{1, 2, 3\}$, for every unit in the sample. The left panel presents treatment 1 ($W = 1$), the middle panel presents treatment 2 ($W=2$), and the right panel presents treatment 3 ($W=3$).}
\label{fig:overlap}
\end{center}
\end{figure*}

Distributions of estimated GPSs across the treatments are compared using boxplots. For each overlap scenario, we estimated the GPS for each unit in the sample using GBM, and plotted the distributions of estimated GPSs using a separate boxplot for the unit receiving each type of treatment (Figure~\ref{fig:overlap}). Substantial overlap in boxplots is presented in the strong overlap scenario, while the weak overlap scenario highlights the different distributions of GPSs.

\subsection{Simulation Results}

\subsubsection{Simulation 1}
Table~\ref{simulation1} displays the MAB, RMSE and CP of the estimates of two ATT effects $ATT_{1|1,2}$ and $ATT_{1|1,3}$, and three ATE effects $ATE_{1,2}$,$ATE_{1,3}$ and $ATE_{2,3}$, for the three scenarios in Simulation 1. 

No single method trumped others in estimating both $ATT_{1|1,2}$ and $ATT_{1|1,3}$ across all three scenarios. For $ATT_{1|1,2}$, outcome modeling approaches had smaller MABs and RMSEs, whereas for $ATT_{1|1,3}$, GPS approaches showed similar or slightly better performance than BART. RA performed best under the scenario of equal sample sizes. As the sample sizes in the comparison groups grew relative to the reference group, BART generally produced low MAB and RMSE. With GPS approaches, IPTW-GBM outperformed IPTW-MLR, IPTW-MLR-Trim, IPTW-SL and IPTW-SL-Trim in the estimates of $ATT_{1|1,2}$ across all three scenarios, but had similar performances in estimating $ATT_{1|1,3}$. Weight trimming did not improve IPTW-MLR, IPTW-GBM or IPTW-SL. VM presented larger bias and RMSE than BART and IPTW-GBM. None of the methods had nominal CP. IPTW methods and RA in general generated greater than the nominal CP, VM had a CP that decreased as the ratio of units became more unbalanced (0.99 to 0.80), and BART yielded a CP around 0.80 -- 0.88, which we suspect is because the reference group is relatively small. Overall, BART and IPTW-GBM tended to show the best performances across settings for the ATT estimates. 

For the ATE estimates, BART consistently provided lower MAB and RMSE  followed by TMLE, across all three scenarios with different ratio of units. BART had nominal CP across all three scenarios. IPTW methods and TMLE yielded conservative intervals and greater than the nominal CP. RA was sensitive to the ratio of units. In the scenario with highly unequal sample sizes across treatment groups, RA had subpar performance. The intervals produced by RA rarely covered the true effects, resulting in a low CP. Altogether, BART and TMLE provided the best performances across settings for the ATE estimates. Boxplots of biases from 200 replications in pairwise ATT and ATE estimates appear in Figure S1 and Figure S2 in Supplemental Materials. 

In Figure~\ref{fig:convergence}, we examined the large-sample convergence property of each of six methods. We considered only the scenario with the ratio of units = 1:15:13, which is the
most representative of the SEER-Medicare registry. We simulated the data with increasing sample sizes of $n = $(2900, 5800, 8700, 11,600, 14,500, 17,400).We computed
the RMSE of the estimates of $ATT_{1|1,2}$ and $ATT_{1|1,3}$ for each $n$. We then regressed $\log(\text{RMSE})$ on ($-\log n$) using a simple linear regression with a slope $b$ for each method. The least-squares estimation of $b$ approximates the convergence rate\cite{liu2013optimal}. BART and GBM converged at a rate of $O(n^{-1/2})$ for both ATT estimates. IPTW-MLR, IPTW-SL, VM and RA all  converged at a slower rate than  $O(n^{-1/2})$. Figure S3 in Supplemental Materials displays the convergence property of each of six method for the estimates of the ATE estimates. BART and TMLE converged at a rate of $O(n^{-1/2})$ for all of the pairwise ATE estimates. GBM varied in the rate of convergence across three pairwise ATE effects, from $O(n^{-1/2})$ to $O(n^{-2/5})$ to $O(n^{-1/3})$.   IPTW-MLR, IPTW-S and RA all had a much slower convergence rate.

\begin{table}[t]
\centering
\footnotesize
\caption{Comparison of the estimated average treatment effects on the treated in terms of mean absolute bias (MAB), root mean square error (RMSE) and coverage probability (CP) across 200 replications in Simulation 1. The causal estimand is based on risk difference.}
\begin{tabular}{clc@{\hspace{1mm}}c@{\hspace{1mm}}c@{\hspace{1mm}}c@{\hspace{1mm}}c@{\hspace{1mm}}c@{\hspace{1mm}}c@{\hspace{1mm}}c@{\hspace{1mm}}c@{\hspace{1mm}}c@{\hspace{1mm}}c@{\hspace{1mm}}c@{\hspace{1mm}}c@{\hspace{1mm}}c@{\hspace{1mm}}c@{\hspace{1mm}}c@{\hspace{1mm}}c@{\hspace{1mm}}c@{\hspace{1mm}}c@{\hspace{1mm}}c}
  \hline
 & &  \multicolumn{3}{c}{$ATT_{1|1,2}$} && \multicolumn{3}{c}{$ATT_{1|1,3}$}&& \multicolumn{3}{c}{$ATE_{1,2}$}&& \multicolumn{3}{c}{$ATE_{1,3}$}&& \multicolumn{3}{c}{$ATE_{2,3}$}\\
   \cline{3-5} \cline{7-9}\cline{11-13}\cline{15-17}\cline{19-21}
  Scenario & Method & MAB & RMSE&CP && MAB & RMSE&CP&& MAB & RMSE&CP&& MAB & RMSE&CP&& MAB & RMSE&CP\\
  \hline
&RA                            & 0.01 & 0.02 &0.99&& 0.03 & 0.04&0.99 && 0.01&0.02&0.99&&0.03&0.04  &0.98 &&0.02&0.02&0.99  \\ 
&IPTW-MLR               & 0.06 & 0.07 &1&& 0.04 & 0.05&1&&0.07&0.08&1&&0.04&0.05&1&&0.09&0.10&1 \\ 
& IPTW-MLR-Trim & 0.06 & 0.07&1 && 0.04 & 0.05&1&&0.07&0.08&1&&0.04&0.05&1&&0.09&0.10&1 \\ 
 &IPTW-GBM              & 0.05 & 0.06 &0.99&& 0.05 & 0.06&0.98&&0.07&0.07&1 &&0.06&0.07&0.98&&0.13&0.13&0.96 \\ 
I&IPTW-GBM-Trim   & 0.06 & 0.07 &0.99&& 0.05 & 0.06&0.98&&0.06&0.06&0.98&&0.06&0.07&1&&0.11&0.12&0.96 \\ 
&IPTW-SL              & 0.06 & 0.07 &1&& 0.05 & 0.06&1&&0.07&0.08&1&& 0.05&0.06&1&&0.12&0.13&1\\ 
&IPTW-SL-Trim              & 0.06 & 0.07 &1&& 0.06 & 0.08&1 &&0.06&0.07&1&&0.05&0.05&1&&0.10&0.11&1\\ 
&VM                & 0.05 & 0.07 &0.99&& 0.06 & 0.08&0.93&&--&--&--&&--&--&--&&--&--&-- \\ 
&BART                     & 0.03 & 0.04&0.88 && 0.04 & 0.05&0.80&&0.03&0.03&0.96&&0.03&0.03&0.95&& 0.03&0.04&0.95\\
&TMLE   &--&--&--&&--&--&--&&0.04&0.05&1&&0.02&0.03&1&&0.05&0.06&1\\
  \hline
&RA                          & 0.02 & 0.02&1 && 0.05 & 0.05&0.92&&0.02&0.02&0.80&&0.05&0.05&0.60&&0.03&0.03&0.67 \\ 
&IPTW-MLR             & 0.05 & 0.06&1 && 0.03 & 0.03&0.99&&0.06&0.08&1&&0.04&0.05&1&&0.07&0.07&1 \\ 
& IPTW-MLR-Trim & 0.06 & 0.06 &1&& 0.03 & 0.03&0.99&&0.06&0.07&1&&0.03&0.04&1&&0.08&0.08&1 \\ 
 &IPTW-GBM            & 0.03 & 0.04&0.98 && 0.03 & 0.04&0.99&&0.05&0.05&0.98&&0.05&0.06&1&&0.09&0.09&0.94 \\ 
II&IPTW-GBM-Trim   & 0.05 & 0.06&0.98 && 0.04 & 0.04&0.99 &&0.05&0.05&0.98 &&0.05&0.06 &1&&0.09&0.09&1 \\ 
&IPTW-SL              & 0.06 & 0.06 &1&& 0.04 & 0.04&0.99 &&0.06&0.07&1&&0.05&0.05&1&&0.11&0.11&1\\ 
&IPTW-SL-Trim              & 0.06 & 0.07&1 && 0.06 & 0.06 &0.99 &&0.06&0.07&1&&0.05&0.05&1&&0.10&0.10&1\\ 
&VM                & 0.04 & 0.05&0.86 && 0.05 & 0.07&0.88&&--&--&--&&--&--&--&&--&--&-- \\ 
&BART                     & 0.02 & 0.03 &0.80&& 0.03 & 0.04 &0.75 &&0.02&0.02&0.96 &&0.01&0.02&0.98 &&0.01&0.02&0.94\\ 
&TMLE  &--&--&--&&--&--&--&&0.04&0.04&1&&0.02&0.02&1&&0.04&0.04&0.96\\
  \hline
&RA                           & 0.03 & 0.03&1 && 0.07 & 0.07&0.44 &&0.03&0.03&0.06 &&0.07&0.07&0.03&&0.04&0.04&0.03\\ 
&IPTW-MLR              & 0.06 & 0.06&1 && 0.02 & 0.03&0.73 &&0.07&0.08&1 &&0.05&0.06& 1&&0.07&0.07&1\\ 
& IPTW-MLR-Trim & 0.06 & 0.07 &1&& 0.02 & 0.03&1 &&0.06&0.07&1&&0.03&0.04&1&&0.07&0.08&1 \\ 
 &IPTW-GBM              & 0.03 & 0.04&1 && 0.02 & 0.03&0.98&&0.04&0.05&0.98 &&0.04&0.05&1 &&0.06&0.06&0.98 \\ 
III&IPTW-GBM-Trim   & 0.06 & 0.06&1 && 0.02 & 0.03&0.98&&0.04&0.05&1&&0.05&0.05&1&&0.06&0.06&0.10 \\ 
&IPTW-SL              & 0.06 & 0.06 &0.99 && 0.03 & 0.03 &1&&0.06&0.07&1 &&0.04&0.05&1 &&0.10&0.10&1\\ 
&IPTW-SL-Trim              & 0.06 & 0.07&0.99 && 0.04 & 0.05&1&& 0.06&0.07&1&&0.04&0.05&1&&0.10&0.10&0.99\\ 
&VM                & 0.03 & 0.04&0.80 && 0.05 & 0.06 &0.78&&--&--&--&&--&--&--&&--&--&--\\ 
&BART                     & 0.02 & 0.03 &0.76&& 0.03 & 0.04&0.74 &&0.02&0.03&0.95 &&0.02&0.03&0.96 &&0.01&0.01&0.94\\  
&TMLE   &--&--&--&&--&--&--&&0.03&0.03&0.98&&0.01&0.02&0.97&&0.03&0.03&0.96\\
   \hline
\end{tabular}
\label{simulation1}
\end{table}

\begin{figure}[ht]
\centering
\includegraphics[scale=0.25]{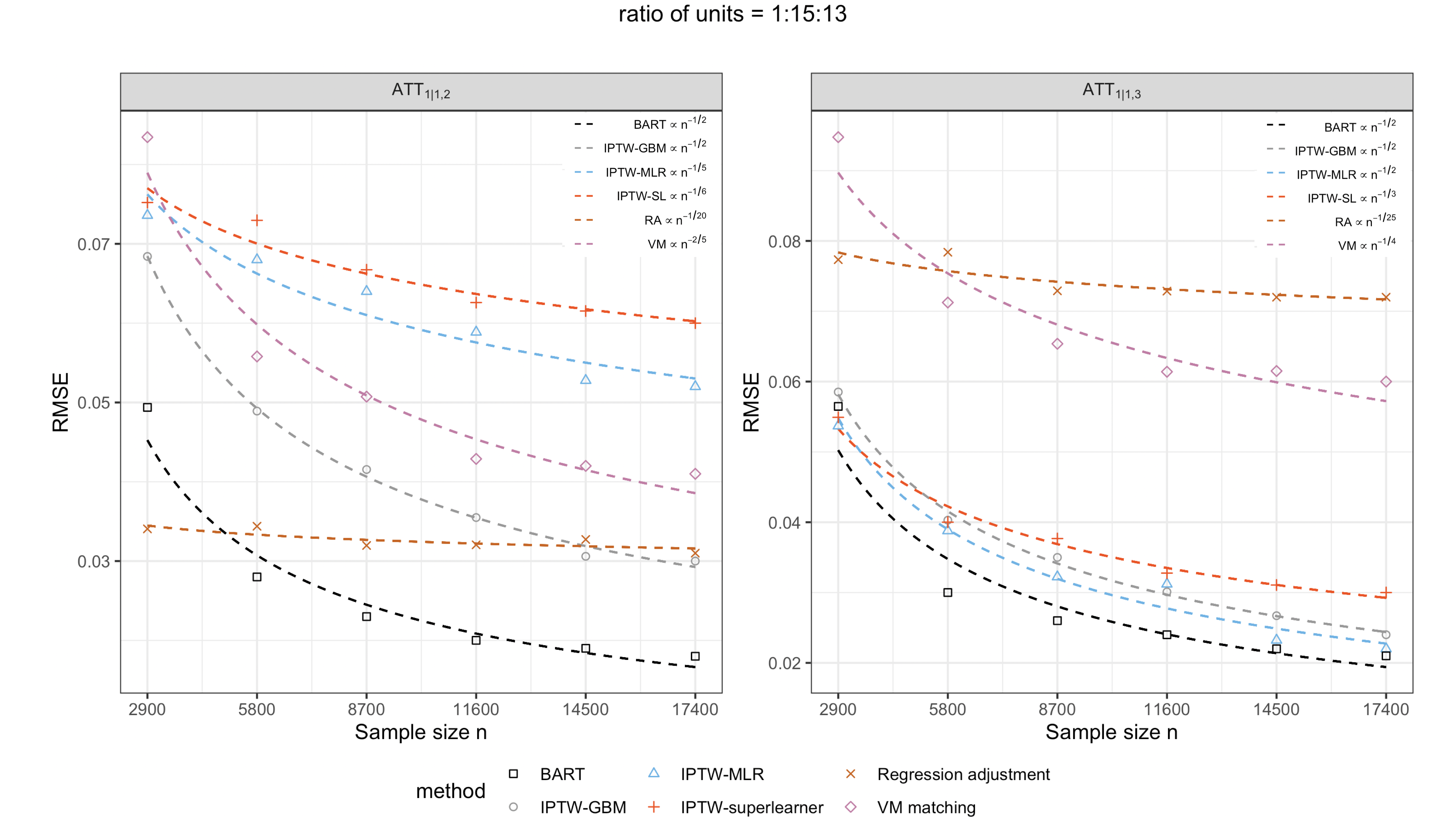}
\caption{The large-sample convergence rate of each of six methods for the estimates of two treatment effects, $ATT_{1|1,2}$ and $ATT_{1|1,3}$. BART and IPTW-GBM converged the fastest, approximately at a rate of $O(n^{-1/2})$. RA converged the slowest,  approximately at a rate of  $O(n^{-1/20})$. }
\label{fig:convergence}
\end{figure}

\subsubsection{Simulation 2}

\begin{figure*}[ht]
\begin{center}
\subfigure[BART-discard vs. GBM for ATT estimates]{\label{fig:disc-att}\includegraphics[scale=0.15]{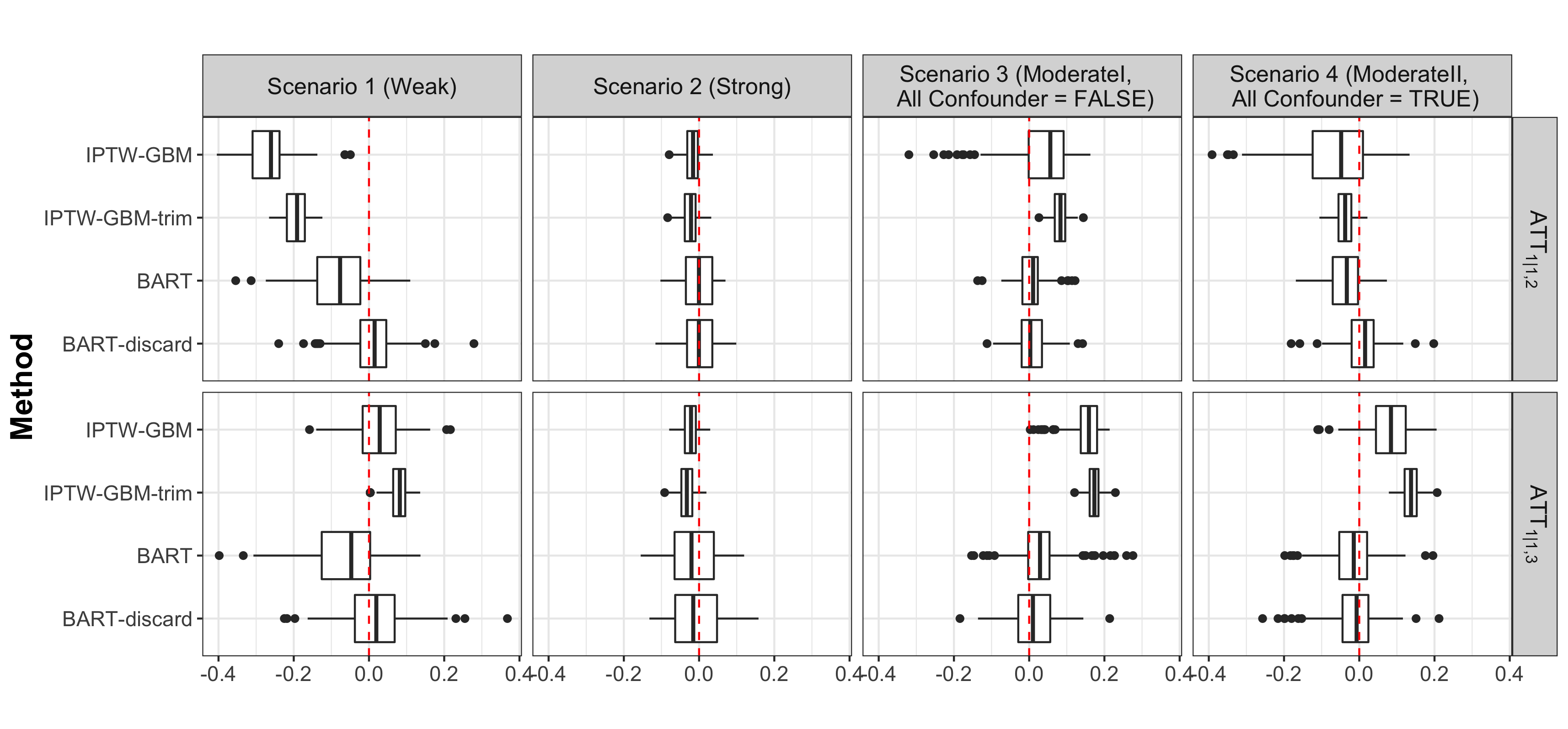}}
\subfigure[BART-discard vs. TMLE for ATE estimates]{\label{fig:disc-ate}\includegraphics[scale=0.15]{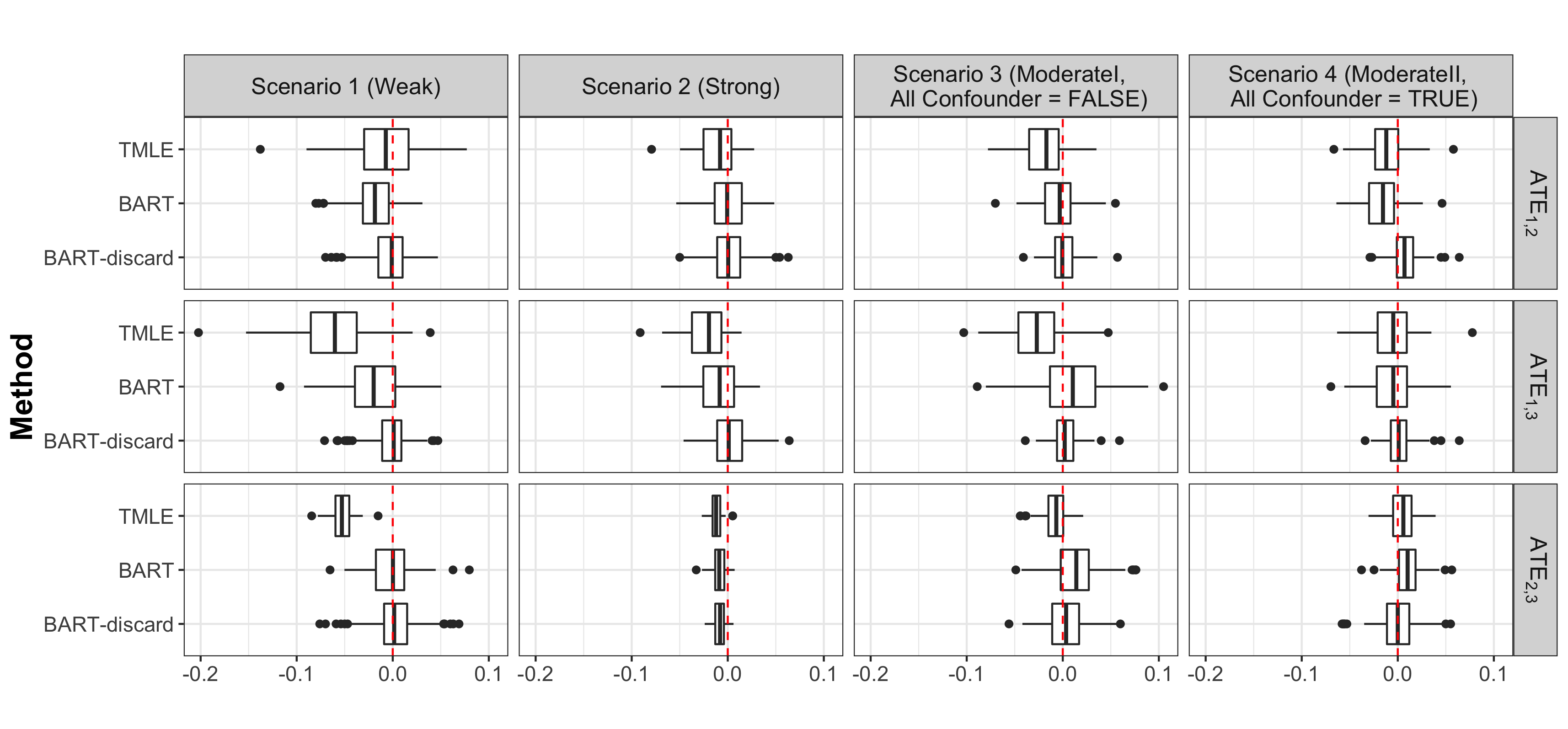}}
\caption{Biases among 200 replications under scenarios of differing covariate overlap for IPTW-GBM vs. BART and two treatment effects $ATT_{1|1,2}$ and $ATT_{1|1,3}$; and for TMLE vs. BART and three treatment effects $ATE_{1,2}$, $ATE_{1,3}$ and $ATE_{2,3}$. 
}
\label{fig:discard}
\end{center}
\end{figure*}

Figure~\ref{fig:discard} displays boxplots of biases of $ATT_{1|1,2}$ and $ATT_{1|1,3}$  among 200 simulations under four levels of overlap for each of IPTW-GBM, IPTW-GBM with trimmed weights, BART and BART with discarding rules (Figure~\ref{fig:disc-att}); and boxplots of biases of $ATE_{1,2}$, $ATE_{1,3}$ and $ATE_{2,3}$ for each of TMLE, BART and BART with discarding rules (Figure~\ref{fig:disc-ate}). 

BART boasts smaller bias under nearly all levels of overlap compared to TMLE and IPTW-GBM. The advantage is more evident when there is more lack of covariate overlap.  The larger biases and RMSEs (see Table S2 in Supplemental Materials) in the IPTW-GBM estimates under the \emph{weak} scenario relative to \emph{moderate} and \emph{strong} overlap suggest that weighting by the GPS -- even by employing flexible machine learning techniques -- suffers from insufficient covariate overlap. The doubly robust method, TMLE, did not show as much variation in its performance across different levels of covariate overlap. In addition, in the \emph{weak} scenario, weight trimming largely altered the IPTW-GBM estimates, indicating the lack of overlap may have led to extreme GPSs. GPS methods ignore the information in the outcome variable, thus assessing covariate overlap regardless of whether the variables are true confounders; BART, on the contrary, takes advantage of the information contributed by the outcome. This is demonstrated by the similar performance delivered by IPTW-GBM in \emph{moderate}I (lack of overlap in non-confounders) and \emph{moderate}II (lack of overlap in true confounders), and better performance of BART in \emph{moderate}I than in \emph{moderate}II. BART perhaps recognized, in \emph{moderate}I, that $X_2, X_3$ and $X_4$ do not play an important role in the response surface, and showed a better performance than IPTW-GBM (smaller bias in both treatment effects).

%Finally, both methods did not necessarily improve as the covariate overlap increased, indicating greater overlap does not necessarily lead to better modeling of the treatment assignment or the outcome generating mechanisms.  Similar evidence was found in examples of McCaffrey et al. \cite{mccaffrey2013tutorial}, where tuning the GBM model to achieve greater balance in the covariates resulted in lower levels of covariate overlap. 
Our BART discarding rule~\eqref{eq:1sd}  considerably reduced the biases in the estimates of both ATE and ATT effects in the \emph{weak} scenario where there was substantial lack of covariate overlap. When the lack of covariate overlap was moderate, the discarding strategy 
noticeably improved over plain BART.  When there was sufficient covariate overlap, BART with and without discarding  performed equally well. The weighting methods and TMLE are not coupled with discarding rules. To get a sense of the proportion of units that would be retained in the common support region for inference based on the GPSs, we applied the GPS-based discarding rule, employed by VM, designed for obtaining a common support region for multiple treatments~\eqref{eq:gpscomsup}. Using BART, the percentages of discarded units in the treated group, averaged across 200 replications, in the \emph{weak}, \emph{moderate}I, \emph{moderate}II and \emph{strong} scenario were  38\%, 24\%, 15\% and 0.2\%,
respectively, as compared to 86\%, 42\%, 42\% and 13\% computed by the GPS-based discarding rule~\eqref{eq:gpscomsup}. BART retains a much larger common support region while providing more accurate treatment effect estimates.

%We expanded Simulation 2 to include results from SBC methods and the methods with multiple treatment comparisons, and highlighted the differences (results in Supplemental Materials). Using SBC significantly biased effect estimates. 

%%%%%%%%%%%%%%%%%%%%%%%%%%%%%%%%%%%%%%%%%%%
\section{Application to SEER-Medicare Data on NSCLC}\label{sec:application}

Clinical encounter and Medicare claims data on 11,980 patients with stage I-IIIA NSCLC was drawn from the latest SEER-Medicare database. These patients were above 65 years of age, diagnosed between 2008 (first year patients in the registry underwent robotic-assisted surgery) and 2013, and underwent surgical resection via one of the three approaches, including robotic-assisted surgery, VATS or open thoractomy. The dataset contains individual-level information at baseline on the following variables: age, gender, marital status, race, ethnicity, income level, comorbidities, cancer stage, tumor size, tumor site, cancer histology and whether they underwent positron emission tomography (PET), chest computer tomography (CT) or mediastinoscopy. Table \ref{Tab.Seer} summarizes these variables for each surgical approach.  We compared the effectiveness of the three surgical approaches in terms of four outcomes: the presence of respiratory complication  within 30 days of surgery or during the hospitalization in which the primary surgical procedure was performed, prolonged length of stay (LOS) (i.e., $>$ 14 days), intensive care unit (ICU) stay following surgery and readmission within 30 days of surgery. Table~\ref{Tab.EventRate} displays the outcome rates in the three surgical groups.

\begin{table*}[t]
\centering
\footnotesize
%\scriptsize
\caption{Baseline characteristics of patients in three surgical groups in SEER-Medicare data.}\label{Tab.Seer}
\begin{tabular}{lccc}
  \hline
 & Robotic-Assisted Surgery & VATS & Open Thoracotomy \\ 
 Characteristics & $N = 396$ & $N = 6582$ & $N = 5002$ \\
 \hline
 Age (years), mean (SD)  & 74.3 (5.7) & 73.9 (5.4) & 74.5 (5.7) \\ 
 Female, N ($\%$) & 223 (56.3) & 3446 (52.4) & 2941 (58.8) \\ 
 Married, N ($\%$) & 227 (57.3) & 3753 (57.0) & 2802 (56.0) \\ 
 Race, N ($\%$) &&&\\
 \;\;White & 320 (80.8) & 5694 (86.5) & 4369 (87.3) \\ 
 \;\;Black & 21 (5.3) & 364 (5.5) & 248 (5.0) \\ 
 \;\;Hispanic & 15 (3.8) & 218 (3.3) & 139 (2.8) \\ 
 \;\;Other & 40 (10.1) & 306 (4.6) & 246 (4.9) \\ 
 Median household annual income, N ($\%$) &&&\\
 \;\;1st quartile & 97 (24.5) & 2132 (32.4) & 1009 (20.2) \\ 
 \;\;2nd quartile & 88 (22.2) & 1729 (26.3) & 1193 (23.9) \\ 
 \;\;3rd quartile & 98 (24.7) & 1345 (20.4) & 1143 (22.9) \\ 
 \;\;4th quartile & 113 (28.5) & 1376 (20.9) & 1657 (33.1) \\ 
Charlson comorbidity score, N ($\%$) &&&\\
 \;\;$0-1$ & 154 (38.9) & 2163 (32.9) & 1810 (36.2) \\ 
 \;\;$1-2$ & 113 (28.5) & 1944 (29.5) & 1379 (27.6) \\ 
 \;\;$>2$  & 129 (32.6) & 2475 (37.6) & 1813 (36.2) \\ 
 Year of diagnosis, N ($\%$) &&&\\
%\;\; 2008 & 2 (0.5) & 1383 (21.0) & 699 (14.0) \\
%\;\; 2009 & 12 (3.0) & 1303 (19.8) & 785 (15.7) \\
\;\;2008-2009 &14(3.5) & 2686 (40.8) & 1484 (29.7)\\
\;\; 2010 & 33 (8.3) & 1123 (17.1) & 857 (17.1) \\
\;\; 2011 & 85 (21.5) & 1033 (15.7) & 866 (17.3) \\
\;\; 2012 & 131 (33.1) & 899 (13.7) & 821 (16.4) \\
\;\; 2013 & 133 (33.6) & 841 (12.8) & 974 (19.5) \\
 Cancer stage, N ($\%$) &&&\\
 \;\;Stage I & 295 (74.5) & 4195 (63.7) & 3884 (77.6) \\ 
 \;\;Stage II & 63 (15.9) & 1504 (22.9) & 709 (14.2) \\ 
 \;\;Stage IIIA & 38 (9.6) & 883 (13.4) & 409 (8.2) \\ 
 Tumor size, in mm, N ($\%$) &&&\\
 \;\;$\leq 20$ & 160 (40.4) & 1967 (29.9) & 2232 (44.6) \\ 
 \;\;$21-30$ & 98 (24.7) & 1696 (25.8) & 1388 (27.7) \\ 
 \;\;$31-50$ & 109 (27.5) & 1804 (27.4) & 987 (19.7) \\ 
  \;\;$\geq 51$ & 29 (7.3) & 1084 (16.5) & 367 (7.3) \\ 
 Histology, N ($\%$) &&&\\
 \;\;Adenocarcinoma & 255 (64.4) & 3757 (57.1) & 3348 (66.9) \\ 
 \;\;Squamous cell carcinoma & 107 (27.0) & 2165 (32.9) & 1167 (23.3) \\ 
 %\;\;Large cell carcinoma & 4 (1.0) & 156 (2.4) & 107 (2.1)  \\ 
 \;\;Other histology & 34 (8.6) & 660 (10.0) &  487 (9.7) \\ 
 Tumor site, N ($\%$) &&&\\
 \;\;Upper lobe & 215 (54.3) & 3829 (58.2) & 2859 (57.2) \\  
 \;\;Middle lobe & 27 (6.8) & 308 (4.7) & 335 (6.7) \\
 \;\;Lower lobe & 141 (35.6) & 2195 (33.3) & 1720 (34.4) \\
 \;\;Other site & 13 (3.3) & 250 (3.8) & 88 (1.8) \\  
 PET scan, N ($\%$) & 302 (76.3) & 5004 (76.0) & 3410 (68.2) \\  
 Chest CT, N ($\%$) & 263 (66.4) & 4525 (68.7) & 3148 (62.9) \\ 
 Mediastinoscopy, N ($\%$) & 62 (15.7) & 715 (10.9) & 420 (8.4) \\
   \hline
\end{tabular}
\footnotesize \\\smallskip
Abbreviations: PET = positron emission tomography; SD = standard deviation; CT = computer tomography
\end{table*}

Among the 11,980 patients,  396 (3.3$\%$) received  robotic-assisted surgery, 6582 (54.9$\%$) underwent VATS, and 5002 (41.8$\%$) were operated via open thoracotomy.   We estimated the causal effects of robotic-assisted surgery vs. VATS or open thoracotomy among patients underwent robotic-assisted surgery (i.e., $ATT_{s_1|s_1,s_2}$ and $ATT_{s_1|s_1,s_3}$) and in the overall population (i.e., $ATE_{s_1,s_2}$ and $ATE_{s_1,s_3}$) using BART, regression adjustment, IPTW with GPSs estimated using multinomial logistic regression or GBM (with or without trimming), and VM. Each method was implemented as described in the simulation section. All pre-treatment covariates were included additively to the GPS models for IPTW methods and VM, and to the response surface models for RA and BART. 

\begin{table*}[t]
\begin{center}
\footnotesize
\caption{The outcome rates in three surgical groups: robotic-assisted surgery, VATS and open thoracotomy.}\label{Tab.EventRate}
\begin{tabular}{l cccc}
  Outcomes& Robotic-Assisted Surgery & VATS & Open Thoracotomy &Overall\\
  & $N = 396$ & $N = 6582$ & $N = 5002$ &$N = 11960$\\\hline
 Respiratory complication& 30.1\% &33.6\% & 33.3\%&33.3\%\\
 Prolonged LOS &5.3\%&10.4\%&5.5\%&8.2\%\\
 ICU Stay &60.2\%&75.3\% &59.1\%&67.9\%\\
 Readmission &8.8\%&9.8\%&8.0\%&9.0\%\\\hline
\end{tabular}\label{tab:eventrate}
\end{center}
\end{table*}

Table S3 in Supplemental Materials presents the point and interval estimates of $ATT_{s_1|s_1,s_2}$ and $ATT_{s_1|s_1,s_3}$ based on risk difference for all the methods examined. To provide uncertainty intervals for the treatment effect estimates, nonparametric bootstrap was used for the IPTW methods and VM, and Bayesian posterior intervals were used for RA and BART. All methods yielded statistically insignificant effects on respiratory complication and readmission if patients who received robotic-assisted surgery had instead been treated with open thoracotomy or VATS. For prolonged LOS and ICU stay, all methods except RA and VM suggested that robotic-assisted surgery led to significant smaller rates of the outcomes compared to open thoracotomy, but  no statistically significant differences compared to VATS. The results from this empirical dataset provided partial evidence that robotic-assisted surgery may have  a positive effect on some postoperative outcomes among those who were operated with robotic-assisted surgery compared to open resection, but no advantages on over VATS. 

To highlight the importance of simultaneous comparisons of multiple treatments, we implemented each method using SBC to show how such inappropriate practices can result in different and confusing estimates of treatment effects. Table S3 also includes the estimates of $ATT_{s_1|s_1,s_2}$ and $ATT_{s_1|s_1,s_3}$ from SBC. For BART, the conclusions are generally consistent with those using multiple treatment comparisons, though we note several inconsistent directions of the estimates of treatment effects. Given the different estimands and subpopulations to which inference using SBC is generalizable when using GPS-based approaches, it would generally be inappropriate to directly compare causal estimates. However, we note that IPTW methods, implemented using SBC, did not always match the findings that were based on IPTW methods designed for multiple treatments. Details appear in Table S4 in Supplementary Materials.

We further explored the sensitivity of BART for binary outcomes to the choice of end-node prior, specifically via the hyperparameter $k$ \citep{dorie2016flexible}. We employed 5-fold cross-validation to choose the optimal $k$ that minimizes the misclassification error. Results suggested the optimal hyperparameter $k = 2$, which is the default value of $k$ in the \texttt{bart()} function (not shown). Moreover, we extended the \textit{1 sd rule}, the discarding rule of BART proposed by Hill and Su \cite{hill2013assessing}, to the multiple treatment setting, to assess whether common support between treatment groups is reasonable based on the uncertainty in the posterior predictive distributions associated with the outcome in the observed versus the counter-factual treatment group. We did not exclude any patients from the empirical dataset based on the discarding rule in~\eqref{eq:1sd}.

%%%%%%%%%%%%%%%%%%%%%%%%%%%%%%%%%%%%%%%%%%%
\section{Summary and Discussion}

Our paper makes two primary contributions to the causal inference literature. First, we extend BART to the multiple treatment and binary outcome setting, highlighting that the strengths of BART for binary treatment also manifest with multiple treatments. Second, we propose a common support rule for BART, and find that BART consistently shows superior performance over alternative approaches in various scenarios with differing levels of covariate overlap. 

In addition to the primary findings in our simulations corresponding to bias, RMSE, CP and large-sample convergence property. BART boasts a few additional advantages that make it a unique tool for the multiple treatment setting. As one example, BART is computationally efficient. All simulations were run in \verb+R+ on a iMAC with a 4GHz Intel Core i7 processor. On a dataset of size $n$=11,600, each BART implementation took less than 150 seconds to run, while each IPTW-GBM implentation took about 10 minutes to run. As a second example, BART produces coherent interval estimates of the treatment effects for either continuous or binary outcomes using posterior samples. For GBM, McCaffrey et al. \cite{mccaffrey2013tutorial} estimate the variance by using robust procedure for continuous outcomes, but acknowledge that there is currently lack of theory to guarantee that this approach results in proper confidence intervals. For estimands based on a binary outcome such as the risk difference investigated in this article, it is difficult to approximate the variance using robust procedure. For matching based approaches, there is still ambiguity regarding appropriate methods for interval estimation \citep{imai2004causal,hill2011bayesian, lopez2017estimation}. 

We apply the methods examined to 11,980 stage I-IIIA NSCLC patients, drawn from the latest SEER-Medicare linkage. Results suggest that robotic-assisted  surgery may be preferred in terms of prolonged LOS and ICU stay, among those who were operated via the robotic-assisted technology, relative to open thoracotomy or VATS. Different choice of methods, or inappropriate practice such as implementing SBC for pairwise ATT effects, may lead to different conclusions about the treatment effects, explicating the importance of appropriate methods and practice for causal inference with multiple treatments. 

The promising performance of BART in the complex multiple treatment settings will lay groundwork for several future research avenues. First, the flexibility offered by nonparametric modeling of BART can be leveraged to model regression relationships in survival data. Second, individual treatment effects that are easily obtained from BART provide a building block for estimating the heterogeneous treatment effect. Finally, we have made a significant untestable assumption related to unmeasured confounding. Developing sensitivity analyses under this complex multiple treatments setting leveraging BART would also be a worthwhile and important contribution.

%\subsection{End of paper special sections}
%Depending on the requirements of the journal that you are submitting to,there are macros defined to typeset various special sections.

\begin{acks}
This study used the linked SEER-Medicare database. The interpretation and reporting of these data are the sole responsibility of the authors. The authors acknowledge the efforts of the National Cancer Institute; the Office of Research, Development and Information, CMS; Information Management Services (IMS), Inc.; and the Surveillance, Epidemiology, and End Results (SEER) Program tumor registries in the creation of the SEER-Medicare database.
\end{acks}

\begin{funding}
 This work was supported in part  by award ME\_2017C3\_9041 from the Patient-Centered Outcomes Research Institute, and by grant P30CA196521-01 from the National Cancer Institute. 
\end{funding}

%\begin{verbatim}
%\begin{sm}
%To typeset a
%  "Supplemental material" section.
%\end{sm}
%\end{verbatim}

\bibliographystyle{SageV} %to use SMMR style
\bibliography{MultiTrts_bart_manuscript}

\end{document}